\documentclass[final,sort&compress]{aipproc}

\layoutstyle{8x11single}

\usepackage{amsmath,wrapfig,bm}

\newif\ifdraft
\ifdraft
\def\note#1{\footnote{\textcolor{red}{#1}}}

\usepackage{svn}
\SVN $Id: AAC.tex 281 2014-10-13 14:22:45Z astamm $
\def\versiontag{\thanks{Version: \SVNId}}
\usepackage{color}

\usepackage{showlabels}

\else
\def\note#1{}
\let\versiontag\relax
\fi
\usepackage{xfrac}
\allowdisplaybreaks

\def\LL{\mathcal{L}}
\def\Lkin{\LL_{\rm part}}
\def\Lint{\LL_{\rm int}}

\def\Lion{\LL_{\rm ion}}
\def\Lfield{\LL_{\rm field}}

\def\V{\varphi}
\def\Vd{\skew{7}\dot \varphi}

\def\ma{m}

\def\qe{q}

\def\wa{w_\alpha}

\def\gammaa{\gamma_\alpha}
\def\gammaad{{\dot\gamma}_\alpha}

\def\qi{q_{\scriptscriptstyle\textsc{I}}}
\def\nion{n^{\scriptstyle\textsc{(Ion)}}}

\def\energy{W}

\def\Np{N_p}

\def\Nz{N_z}
\def\Nr{N_r}

\def\sump{\sum_{\alpha=1}^{\Np}}

\def\sumrl{\sum_{l=1}^{\Nr}}
\def\sumzk{\sum_{k=1}^{Nz}}
\def\sumra{\sum_{a=1}^{\Nr}}
\def\sumza{\sum_{a=1}^{Nz}}
\def\sumrb{\sum_{b=1}^{\Nr}}
\def\sumzb{\sum_{b=1}^{Nz}}

\def\D#1,#2;{D^{\scriptscriptstyle(#1)}_{#2}}

\def\xid{\dot{\boldsymbol \xi}}

\def\xia{{\boldsymbol \xi}^\alpha}

\def\xida{\xid^\alpha}

\def\pia{{\boldsymbol \pi}^\alpha}

\def\xixd{\dot\xi_x}

\def\xixa{\xi_x^\alpha{}}
\def\xixda{\xixd^\alpha{}}

\def\xiyd{\dot\xi_y}

\def\xiya{\xi_y^\alpha{}}
\def\xiyda{\xiyd^\alpha{}}

\def\xizd{\dot\xi_z}

\def\xiza{\xi_z^\alpha{}}
\def\xizda{\xizd^\alpha{}}

\def\pihd{\dot\pi_h}
\def\piha{\pi_h^\alpha{}}
\def\pihda{\pihd^\alpha{}}

\def \A{{\bf A}}
\def \Avd{\dot{\bf  A}}

\def\x{{\bf x}}
\def\v{{\bf v}}

\def\argxv{\x, \v}

\def\Ad{\skew{3}\dot A}
\def\Add{\ddot A}

\def\dz{\Delta z}
\def\dr{\Delta r}

\def\d3#1{\mathrm{d}^3#1}

\def \Dr{D^{(r)}}
\def \D{D^{(z)}}

\begin{document}

\title{Variational Formulation of E \& M Particle Simulation Algorithms in Cylindrical Geometry using an Angular Modal Decomposition\versiontag}

\author{A. B. Stamm}{address={Department of Physics and Astronomy, University of Nebraska-Lincoln, Lincoln, NE 68588-0299, USA}}
\author{B. A. Shadwick}{address={Department of Physics and Astronomy, University of Nebraska-Lincoln, Lincoln, NE 68588-0299, USA}}

\begin{abstract}
Taking advantage of the flexibility of the variational method with coordinate transformations, we derive a self-consistent set of equations of motion from a discretized Lagrangian to study kinetic
plasmas using a Fourier decomposed cylindrical coordinate system.  The phase-space distribution function was reduced to a collection of finite-sized macro-particles of arbitrary shape moving on a
virtual Cartesian grid.  However, the discretization of field quantities was performed in cylindrical coordinates and decomposed into a truncated Fourier series in angle.  A straightforward finite
element interpolation scheme is used to transform between the two grids.  The equations of motion were then obtained by demanding the action be stationary.  The primary advantage of the variational
approach is preservation of Lagrangian symmetries.  In the present case, this leads to exact energy conservation, thus avoiding possible difficulties with grid heating.
\end{abstract}

\maketitle

\section{Introduction}

Several approaches have been used to reduce the computational costs of 3D Particle-in-Cell (PIC) simulations while maintaining predictive power.  Some examples include working in the moving window
reference frame, using quasi-static approximations \cite{Huang:2006ve,Mehrling:2014aa}, using the ponderomotive guiding center algorithm \cite{Gordon00} and the quasi-3D method
\cite{Lifschitz:2009aa,Davidson:2014aa}.  Lifschitz \textit{et al.}~\cite{Lifschitz:2009aa} exploit the near cylindrical symmetry of the laser-plasma interaction by expanding the electromagnetic
fields in a small number of poloidal modes while retaining the full 3D dynamics of the macro-particles.  By projecting the macro-particle currents onto the $r$--$z$ grid, all macro-particles in a
given annulus contribute current to the same set of grid points, greatly reducing sampling noise.

Recently a variation formulation of macro-particle model has been developed \cite{Evstatiev:2013aa,Shadwick:2014aa,Stamm:2014aa} that inherits all the benefits of a Lagrangian approach, specifically
the connection between symmetries and conservation laws and the flexibility in choosing coordinates.  Here we apply this technique to a incorporate a modal decomposition of the electromagnetic fields
into a macro-particle model.  As in previous work \cite{Evstatiev:2013aa,Shadwick:2014aa,Stamm:2014aa}, we begin with the relativistic Low Lagrangian \cite{Low:1958aa} and introduce a distribution
function consisting of a collection of fixed-size macro-particles of the form
\begin{equation}
	f(\argxv, t) = \sump\wa\,S_x[x-\xixa(t)]\,S_y[y-\xiya(t)]\,S_z[z-\xiza(t)]\,\delta[v_x - \xixda(t)]\delta[v_y - \xiyda(t)]\delta[v_z - \xizda(t)],
\end{equation}
where $\wa$ is the macro-particle weight, $\xia(t)$ and $\xida(t)$ are the macro-particle position and velocity respectively, $S_x$, $S_y$ and $S_z$ are 1D shape functions \cite{Evstatiev:2013aa} and
$\delta$ is the Dirac delta function.  By taking the macro-particle shape to be the product of 1D shape functions, we are implicitly defining a regular Cartesian grid whose parameters are set by the
particle sizes in $x$, $y$, and $z$.  As a result, it proves convenient to express the vector potential in terms of its Cartesian components, $A_x$, $A_y$, and $A_z$, while parameterizing the
functional dependence of these components by $r$, $\theta$, $z$, and $t$.  We then use a truncated Fourier representation in $\theta$ for all the scalar and vector potentials, keeping only the first
three modes:
\begin{equation}
\V(r,\theta,z,t) = \V^{(o)}(r,z,t) + \V^{(c)}(r,z,t) \,\cos\theta + \V^{(s)}(r,z,t)\,\sin\theta
\end{equation}
and likewise for $\A$.  In most circumstances, this adequately describes the wakefield (the dc term) and the laser field (the first order terms) as discussed in Ref.~\cite{Lifschitz:2009aa}.

\section{Discrete Lagrangian}
We introduce uniform radial and longitudinal grids with $\Nr$ grid points and spacing $\dr$ and $\Nz$ grid points and spacing $\dz$, respectively and denote the coordinates of the grid points by
$r_l$, $l\in[1,\Nr]$ and $z_k$, $k\in[1,\Nz]$.  We denote the approximation of the potentials at $r_l$ and $z_k$ by $\V^{(m)}_{lk}(t)$ and $\A^{(m)}_{lk}(t)$ with $m\in\{o, s, c\}$.  The fully
discretized Lagrangian takes the form $\LL = \Lkin + \Lint + \Lfield + \Lion$.  The particle contribution can be readily computed from the form of the distribution function
\cite{Evstatiev:2013aa,Stamm:2014aa}, while the field terms in the Lagrangian can be readily evaluated by representing the spatial derivatives by finite differences:
\begin{flalign}
	\Lkin &= -\ma c^2\sump\wa\,\sqrt{1 - \frac{\big(\xida \big)^2}{c^2}},\\
	\Lfield  &=\frac18 \sumzk\sumrl r_l\Biggl\{
	\Bigg[
	2\left(\frac1c\,\Ad^{(o)}_{x,lk}\right)^2 + \left(\frac1c\,\Ad^{(c)}_{x,lk}\right)^2 + \left(\frac1c\,\Ad^{(s)}_{x,lk}\right)^2
	+2\left(\frac1c\,\Ad^{(o)}_{y,lk}\right)^2 + \left(\frac1c\,\Ad^{(c)}_{y,lk}\right)^2 + \left(\frac1c\,\Ad^{(s)}_{y,lk}\right)^2 
	 \nonumber &\\
	&
	+ \frac2c\,\Ad^{(c)}_{x, lk}\sumra\Dr_{al}\, \V_{ak}^{(o)}
	+ \frac2c\,\Ad^{(o)}_{x, lk}\left(\sumra\Dr_{al}\, \V_{ak}^{(c)} +\frac{1}{r_l} \V_{lk}^{(c)} \right)
	+ \frac2c\,\Ad^{(s)}_{y, lk}\sumra\Dr_{al}\, \V_{ak}^{(o)}	
	+ \frac2c\,\Ad^{(o)}_{y, lk}\left(\sumra\Dr_{al}\, \V_{ak}^{(s)} +\frac{1}{r_l} \V_{lk}^{(s)} \right)\nonumber\\
	& +2\left(\sumra\Dr_{al}\, \V_{ak}^{(o)}\right)^2 
	+\left(\sumra\Dr_{al}\, \V_{ak}^{(c)}\right)^2  
	+\left(\sumra\Dr_{al}\, \V_{ak}^{(s)}\right)^2 
	+ \left(\frac{1}{r_l} \V_{lk}^{(c)} \right)^2 
	+ \left(\frac{1}{r_l} \V_{lk}^{(s)} \right)^2   \nonumber\\
	& +2\left(\sumzb\D_{bk}\V^{(o)}_{lb} +  \frac1c\,\Ad^{(o)}_{z,lk}\right)^2
	+ \left(\sumzb\D_{bk}\V^{(c)}_{lb} + \frac1c\,\Ad^{(c)}_{z,lk}\right)^2
	+ \left(\sumzb\D_{bk}\V^{(s)}_{lb}\frac1c\,\Ad^{(s)}_{z,lk}\right)^2 
	\Bigg] \nonumber \\[2pt] 
 	-&\Bigg[ 
	2\left(\sumzb\D_{bk}\, A^{(o)}_{y, lb}\right)^2 
	+\left(\sumra\Dr_{al}\, A^{(o)}_{z, ak} - \sumzb\D_{bk}\, A^{(s)}_{y, lb} \right)^2 
	- \sumzb\D_{bk}\, A^{(o)}_{y, lb}  \left(\sumra\Dr_{al}\, A^{(s)}_{z, ak} - \frac{1}{r_l} A^{(s)}_{z, lk}\right)
	+ \left(\frac{1}{r_l} A^{(s)}_{z, lk}\right)^2   \nonumber \\
	&+2\left(\sumzb\D_{bk}\, A^{(o)}_{x, lb}\right)^2 
	+\left(\sumra\Dr_{al}\, A^{(o)}_{z, ak} - \sumzb\D_{bk}\, A^{(c)}_{x, lb} \right)^2 
	- \sumzb\D_{bk}\, A^{(o)}_{x, lb}  \left(\sumra\Dr_{al}\, A^{(c)}_{z, ak} - \frac{1}{r_l} A^{(c)}_{z, lk}\right)
	+ \left(\frac{1}{r_l} A^{(c)}_{z, lk}\right)^2   \nonumber \\ 
	&-\frac12\left(\sumra\Dr_{al}\, A^{(c)}_{y, ak} - \frac1{r_l}\, A^{(s)}_{x, lk} \right)\left(\sumra\Dr_{al}\, A^{(s)}_{x, ak} - \frac1{r_l}\, A^{(c)}_{y, lk} \right)	
	+\left(\sumzb\D_{bk}\, A^{(c)}_{y, lb}\right)^2 +\left(\sumzb\D_{bk}\, A^{(s)}_{x, lb}\right)^2 \nonumber \\ 
	&+\left(\sumra\Dr_{al}\, A^{(o)}_{y, ak}\right)^2  
	+\left(\sumra\Dr_{al}\, A^{(o)}_{x, ak}\right)^2 
	+\frac34\left(\sumra\Dr_{al}\, A^{(c)}_{y, ak} - \frac1{r_l}\, A^{(s)}_{x, lk} \right)^2 
	+\frac34\left(\sumra\Dr_{al}\, A^{(s)}_{x, ak} - \frac1{r_l}\, A^{(c)}_{y, lk} \right)^2   \nonumber \\ 
	 &+ \left(\sumra\Dr_{al}\, A^{(s)}_{z, ak}\right)^2  
	 + \left(\sumra\Dr_{al}\, A^{(c)}_{z, ak}\right)^2    
	+\frac14\left(\sumra\Dr_{al}\, A^{(s)}_{y, ak} - \sumra\Dr_{al}\, A^{(c)}_{x, ak} - \frac1{r_l} A^{(s)}_{y, lk} + \frac1{r_l} A^{(c)}_{x, lk}\right)^2 
	\Bigg]\Biggl\},
\end{flalign}
where $\Dr_{ab}$ and $\D_{ab}$ are the finite difference representations of first derivatives in $r$ and $z$ respectively. 

The interaction term, however, requires knowledge of the potentials at arbitrary spatial locations.  The most straightforward approach is to treat the macro-particles as moving on a 3D \textit{virtual}
Cartesian grid.  The potential on this virtual grid are interpolated between the grid points using finite elements \cite{Evstatiev:2013aa}, allowing evaluation of the interaction Lagrangian.  In terms
of the potentials on the virtual grid, the interaction Lagrangian can be written as
\begin{equation}
	\Lint =\qe\sump\wa \sum_{i,j,k}\rho_{ijk}(\xia)\left[\frac1c\,\,\xida\cdot\A_{ijk} - \V_{ijk}\right]\,,
\end{equation}
where $\V_{ijk}$ and $\A_{ijk}$ are the values of the potential on the virtual grid, and $\rho_{ijk}$ is the finite-element projection of 
\\[2pt]
$S_x(x-\xixa)\,S_y(y-\xiya)\,S_z(z-\xiza)$~\cite{Evstatiev:2013aa}.  
Assuming a static ion background, $\Lion$ can be written as
%To simplify loading macro-particles, we interpolate the (fixed) ion density to the virtual grid to compute $\Lion$ as
\begin{equation}
	\Lion = -\qi\sum_{i,j,k}\nion_{ijk}\V_{ijk}\,,
\end{equation}
\\[-6pt]
where the coefficients $\nion_{ijk}$ define the fixed ion charge distribution on the virtual Cartesian grid.

%\\[-10pt]
\begin{wrapfigure}[19]{r}{3in}
	\vspace{-8pt}
	\begin{center}
	\includegraphics[scale=0.9]{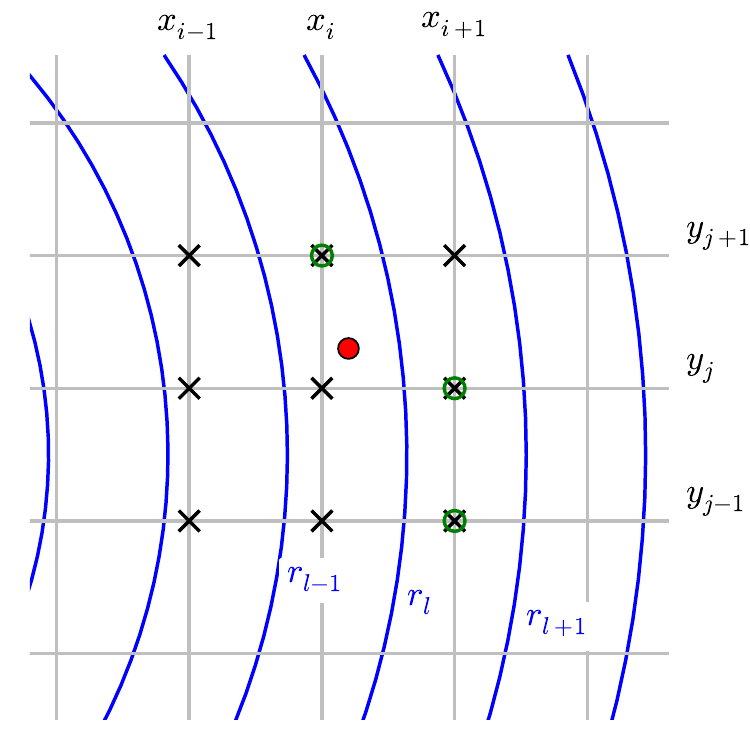}
	\end{center}
	\vskip-12pt
	%% the \caption command doesn't work properly with wrapfig (likely due to stupid tricks with the figure environment, so simluate the caption command
	\hbox to 3in {\hskip 0.125in \parbox{2.85in} {\footnotesize\textbf{FIGURE \ref{grids}.} Planar illustration of the potential interpolation assuming a quadratic particle shape.  For a given
	macro-particle position (red disk), nine locations on the virtual Cartesian grid (gray lines) are involved in determining the potentials at the macro-particle (black crosses).  The potential values on
	the virtual grid are obtained from $\V^{(s)}_{lk}$ and $\A^{(m)}_{lk}$ on the radial grid (blue lines).  For example, using quadratic interpolation in $r$, the potentials at the virtual grid
	points indicated with the green squares are obtained using the radial grid points $l-1$, $l$, and $l+1$.}} \refstepcounter{figure}\label{grids}
\end{wrapfigure}

The potentials on the virtual grid are obtained from the Fourier decomposition along with interpolation in $r$ (see Fig.\ref{grids}):
\begin{multline}
	\V_{ijk} = \sum_{l} \Lambda^{ij}_l \left( \V^{(o)}_{lk} + \V^{(c)}_{lk}\, \cos\theta + \V^{(s)}_{lk}\, \sin\theta \right)\\[-8pt]
	= \sum_{l} \Lambda^{ij}_l \left( \V^{(o)}_{lk} + \frac{x_i}{r_{ij}}\,\V^{(c)}_{lk} + \frac{y_j}{r_{ij}}\,\V^{(s)}_{lk}\right)\,, \label{conversion} 
\end{multline}
where $\Lambda^{ij}_l$ are the interpolation coefficients, $r_{ij}=\sqrt{x_i^2+y_j^2}$ and the $k$ index is simply carried through since the $z$-axis is unaffected by the transformation.  An
analogous expression relates $\A_{ijk}$ to $\A^{(m)}_{ak}$.

\vspace{2pt}
This approach of adopting a virtual grid has general applicability as it decouples the computational grid used to advance the potentials (or fields) from the grid used to interpolate forces. 
For example, in the case of a non-uniform grid, this approach eliminates the need to have the macro-particle size conform to the local grid spacing.

\section{Discrete Equations of Motion}

The equations of motion are obtained by requiring the action to be stationary under variations of the particle position and potentials.  For the particles, the
Euler--Lagrange equations give
\begin{equation}
	\frac{d\pia}{dt} = \frac\qe c \sum_{i,j,k}\left[\left(\xida\cdot\A_{ijk}-c\,\V_{ijk}\right)\frac{\partial\rho_{ijk}(\xia)}{\partial\xia} - \dot\A_{ijk}\,\rho_{ijk}(\xia) -
	\A_{ijk}\,\xida\cdot\frac{\partial\rho_{ijk}(\xia)}{\partial\xia} \right]\,,\label{LorentzF}
\end{equation}
where $\pia \equiv \ma\,\gammaa\,\xida$ is the usual relativistic momentum, with $\gammaa = \sqrt{1 +|\pia|^2/m^2c^2}$.  Variation with respect to $\V^{(m)}_{lk}$ yields Poisson's equation, leading to
\begin{flalign}
	&\sumra  r_a \,\Dr_{la}\left(\frac1c\,\Ad^{(c)}_{x, ak} + \frac1c\,\Ad^{(s)}_{y, ak} + 2\sumrb\Dr_{ba}\V^{(o)}_{bk}\right)
	+ 2\,r_l\sumza\D_{ka}\left(\frac1c\,\Ad^{(o)}_{z, la} + \sumzb\D_{ba}\,\V^{(o)}_{lb}\right)\nonumber \\
	&\hskip240pt = 4\sum_{i,j} \left(\qe\sump\wa\,\rho_{ijk}(\xia) + \qi\,\nion_{ijk} \right) \Lambda_l^{ij}\,,\label{Poisson0}&\\
	&\frac1{r_l}\V^{(c)}_{lk} + \frac1c\,\Ad^{(o)}_{x, lk} + \sumra r_a\,\D_{la}\left(\frac1c\,\Ad^{(o)}_{x, ak} + \sumrb\Dr_{ba}\V^{(c)}_{bk}\right)
	+ r_l\sumza \D_{ka}\left(\frac1c\,\Ad^{(c)}_{z, la} + \sumzb\D_{ba}\,\V^{(c)}_{lb}\right)\nonumber\\
	&\hskip240pt = 4\sum_{i,j}\left(\qe\sump\wa\,\rho_{ijk}(\xia) + \qi\,\nion_{ijk}\right)\Lambda_l^{ij}\,\frac{x_i}{r_{ij}}\,, \label{PoissonC}&\\
	&\frac1{r_l}\V^{(s)}_{lk}+\frac1c\,\Ad^{(o)}_{y, lk} + \sumra r_a\,\D_{la}\left(\frac1c\,\,\Ad^{(o)}_{y, ak} + \sumrb\Dr_{ba}\V^{(s)}_{bk}\right)
	+ r_l\sumza \D_{ka}\left(\frac1c\,\Ad^{(s)}_{z, la} + \sumzb\D_{ba}\,\V^{(s)}_{lb}\right)\nonumber\\
	&\hskip240pt = 4\sum_{i,j}\left(\qe\sump\wa\,\rho_{ijk}(\xia) +\qi\,\nion_{ijk}\right)\Lambda_l^{ij}\,\frac{y_j}{r_{ij}}\,. \label{PoissonS}&
\end{flalign}
Similarly, variation with respect to $\A^{(m)}_{lk}$  leads to the following modal wave equations:
\begin{flalign}
	&\frac{2 r_l}{c^2}\,\Add^{(o)}_{x, lk} + \frac1c\,\,\Vd^{(c)}_{lk} + \frac{ r_l}c \sumra\Dr_{al}\,\Vd^{(c)}_{ak} 
	+ 2\,r_l\sumza\sumzb \D_{ka}\,\D_{ba}\,A^{(o)}_{x, lb} + \sumra\sumrb r_a\,\Dr_{la}\,\Dr_{ba}\,A^{(o)}_{x, bk} \nonumber \\
	&\hskip100pt - \frac12\sumzb\D_{kb}\left(\sumra r_a\,\Dr_{al}\,A^{(c)}_{z, ab} - A^{(c)}_{z, lb}\right) 
	= 4\,\qe\sump\wa\,\sum_{i,j}\rho_{ijk}(\xia)\left(\frac1c\,\xixda\,\Lambda_l^{ij}\right), \label{WaveAxo}\\
	&\frac{2 r_l}{c^2}\,\Add^{(o)}_{y, lk} + \frac1c\,\,\Vd^{(s)}_{lk} + \frac{r_l}c\sumra\Dr_{al}\,\Vd^{(s)}_{ak} 
	+ 2\, r_l\sumza\sumzb\D_{ka}\,\D_{ba}\,A^{(o)}_{x, lb} + \sumra\sumrb r_a\,\Dr_{la}\,\Dr_{ba}\,A^{(o)}_{y, bk}\nonumber \\
	&\hskip100pt -\frac12\sumzb\D_{kb}\left(\sumra r_a\,\Dr_{al}\,A^{(s)}_{z, ab} - A^{(s)}_{z, lb} \right)
	= 4\,\qe\sump\wa\,\sum_{i,j}\rho_{ijk}(\xia) \left(\frac1c\,\xiyda\, \Lambda_l^{ij} \right) \label{WaveAyo}\\
	&\frac{2 r_l}{c^2}\,\Add^{(o)}_{z, lk} + \frac{2 r_l}c\sumzb\D_{bk}\,\Vd^{(o)}_{lb} - \sumra\sumzb r_a\,\Dr_{la}\,\D_{bk}\left(A^{(s)}_{y, ab} + A^{(c)}_{x, ab}\right)
	+ 2\sumra\sumrb r_a\,\Dr_{la}\,\Dr_{ba}\,A^{(o)}_{z, bk} \nonumber \\
	&\hskip100pt = 4\,\qe\sump\wa\,\sum_{i,j}\rho_{ijk}(\xia) \left(\frac1c\,\xizda\, \Lambda_l^{ij} \right), \label{WaveAzo}\\
	&\frac{r_l}{c^2}\,\Add^{(c)}_{x, lk} + \frac{r_l}c\sumra\Dr_{al}\,\Vd^{(o)}_{ak} 
	+ r_l\sumza\sumzb\D_{ka}\,\D_{ba}\,A^{(c)}_{x, lb} - r_l\sumra\sumzb\D_{kb}\,\Dr_{al}\,A^{(o)}_{z, ab}
	+ \frac14\sumra\sumrb r_b\,\Dr_{lb}\,\Dr_{ab}\Big(A^{(c)}_{x, ak} - A^{(s)}_{y, ak}\Big)  \nonumber \\
	&\qquad
	-\frac14\sumra \Big(\Dr_{al} + \Dr_{la}\Big)\Big( A^{(c)}_{x, ak} - A^{(s)}_{y, ak} \Big)
	+\frac1{4\, r_l}\Big(A^{(c)}_{x, lk} -  A^{(s)}_{y, lk} \Big) 
	= 4\,\qe\sump\wa\,\sum_{i,j}\rho_{ijk}(\xia) \left(\frac1c\,\xixda\, \Lambda_l^{ij}\frac{x_i}{r_{ij}} \right),  \label{WaveAxc}\\
	&\frac{r_l}{c^2}\,\Add^{(c)}_{y, lk} + r_l\sumza\sumzb\D_{ka}\,\D_{ba}\,A^{(c)}_{y, lb}
	+ \frac34\, r_l \sumra\sumrb\Dr_{lb}\,\Dr_{ab}\,A^{(c)}_{y, ak} + \frac34\frac1{r_l}\,A^{(c)}_{y, lk} - \frac34\sumra\left(\Dr_{la} +\Dr_{al}\right)A^{(s)}_{x, ak}\nonumber \\
	&\qquad
	- \frac14\sumra\sumrb r_b\,\Dr_{lb}\,\Dr_{ab}\,A^{(s)}_{x, ak} - \frac1{4\, r_l}\,A^{(s)}_{x, lk} + \frac14\sumra\left(\Dr_{la} +\Dr_{al}\right)A^{(c)}_{y, ak} 
	= 4\,\qe\sump\wa\,\sum_{i,j}\rho_{ijk}(\xia) \left(\frac1c\,\xiyda\, \Lambda_l^{ij}\frac{x_i}{r_{ij}} \right),  \label{WaveAyc}\\
	&\frac{r_l}{c^2}\,\Add^{(c)}_{z, lk}  + \frac{r_l}c\sumzb\D_{bk}\,\Vd^{(c)}_{lb} + \sumra\sumrb r_b\,\Dr_{lb}\,\Dr_{ab}\,A^{(c)}_{z, ak} 
	+ \frac1{r_l}\,A^{(c)}_{z, lk} + \sumzb\D_{bk}\,A^{(o)}_{x, lb} - \sumra\sumzb\Dr_{la}\,\D_{bk}\,A^{(o)}_{x, ab}\nonumber \\
	&\hskip100pt
	= 4\,\qe\sump\wa\,\sum_{i,j}\rho_{ijk}(\xia)\left(\frac1c\,\xizda\,\Lambda_l^{ij}\frac{x_i}{r_{ij}} \right),  \label{WaveAzc}\\
	&\frac{r_l}{c^2}\,\Add^{(s)}_{x, lk} + {r_l}\sumza\sumzb\D_{ka}\,\D_{ba}\,A^{(s)}_{x, lb} + \frac34\,r_l\sumra\sumrb\Dr_{lb}\,\Dr_{ab}\,A^{(s)}_{x, ak} 
	+ \frac34\frac1{r_l}\,A^{(s)}_{x, lk} - \frac34\sumra\left(\Dr_{la} + \Dr_{al}\right)A^{(c)}_{y, ak}\nonumber \\
	&\qquad
	- \frac14\sumra\sumrb r_b\,\Dr_{lb}\,\Dr_{ab}\,A^{(c)}_{y, ak} - \frac1{4\, r_l}\,A^{(c)}_{y, lk} + \frac14\sumra\left(\Dr_{la} +\Dr_{al}\right)A^{(s)}_{x, ak} 
	= 4\,\qe\sump\wa\,\sum_{i,j}\rho_{ijk}(\xia) \left(\frac1c\,\xixda\,\Lambda_l^{ij}\frac{y_j}{r_{ij}}\right),  \label{WaveAxs}\\
	&\frac{r_l}{c^2}\,\Add^{(s)}_{y, lk} + \frac{r_l}c\sumra\Dr_{al}\,\Vd^{(o)}_{ak} + r_l\sumza\sumzb\D_{ka}\,\D_{ba}\,A^{(s)}_{y, lb} - r_l\sumra\sumzb\D_{kb}\,\Dr_{al}\,A^{(o)}_{z, ab}
	- \frac14\sumra\sumrb r_b\,\Dr_{lb}\,\Dr_{ab}\Big(A^{(c)}_{x, ak} - A^{(s)}_{y, ak}\Big)  \nonumber \\
	&\qquad
	+\frac14\sumra \Big(\Dr_{al} + \Dr_{la}\Big)\Big( A^{(c)}_{x, ak} - A^{(s)}_{y, ak} \Big)
	-\frac1{4\, r_l}\Big(A^{(c)}_{x, lk} -  A^{(s)}_{y, lk} \Big) 
	= 4\,\qe\sump\wa\,\sum_{i,j}\rho_{ijk}(\xia) \left(\frac1c\,\xiyda\, \Lambda_l^{ij}\frac{y_j}{r_{ij}} \right),  \label{WaveAys}\\
	&\frac{r_l}{c^2}\,\Add^{(s)}_{z, lk}  + \frac{r_l}c\sumzb\D_{bk}\,\Vd^{(s)}_{lb} + \sumra\sumrb r_b\,\Dr_{lb}\,\Dr_{ab}\,A^{(s)}_{z, ak} 
	+ \frac1{r_l}\,\,A^{(s)}_{z, lk} + \sumzb\D_{bk}\,A^{(o)}_{y, lb} - \sumra\sumzb r_a\,\Dr_{la}\,\D_{bk}\,A^{(o)}_{y, ab}\nonumber\\
	&\hskip100pt
	= 4\,\qe\sump\wa\,\sum_{i,j}\rho_{ijk}(\xia) \left(\frac1c\,\xizda\, \Lambda_l^{ij}\frac{y_j}{r_{ij}} \right).  \label{WaveAzs}
\end{flalign}

\section{Energy}

Since the Lagrangian has no explicit time dependence there is a conserved energy of the form
\begin{equation}
	\energy = \sump\xida\cdot\frac{\partial\LL}{\partial\xida} + 
	\sum_{l,k} \left(\dot\A_{lk}^{(o)}\cdot\frac{\partial\LL}{\partial\dot\A_{lk}^{(o)}} + 
	\dot\A_{lk}^{(c)}\cdot\frac{\partial\LL}{\partial\dot\A_{lk}^{(c)}}
	+ \dot\A_{lk}^{(s)}\,\frac{\partial\LL}{\partial\dot\A_{lk}^{(s)}}\right)	-\LL\,.
\end{equation}
Using the equations of motions, it is straightforward to show that this energy is a conserved quantity.  After computing the time derivative of the energy, one can use the Lorentz force equation
\eqref{LorentzF} to evaluate the kinetic energy term through ${\dot\gamma}_\alpha =\sum_h\pihda\,\partial\gammaa/\partial\piha$ giving $mc^2\,\gammaad = -\qe \sum_{i,j,k}\left[\V_{ijk}\,\dot
\rho_{ijk}(\xia) - (1/c)\xida\cdot\Avd_{ijk}\,\rho_{ijk}(\xia)\right]$. The remaining terms in the particle summation can be evaluated by using the modal Poisson and wave equations.  Specifically, to
evaluate the first term in the brackets, we take the time derivative of the modal Poisson equations \eqref{Poisson0}--\eqref{PoissonS} before summing with the corresponding $\V^{(m)}_{lk}$ to make use 
of \eqref{conversion}.  We apply the same approach to obtain the second term summing the modal wave equation \eqref{WaveAxo}--\eqref{WaveAzs} with ${\dot\A}_{lk}^{(m)}$.  This leaves
only field-like terms which cancel to zero, proving that $d \energy/dt =0$.

\section{Conclusion}

Building on previous work \cite{Evstatiev:2013aa,Shadwick:2014aa,Stamm:2014aa}, we have derived a macro-particle model that exploits the near cylindrical symmetry of the laser-plasma interaction by
introducing a poloidal mode expansion of the potentials.  We have obtained a set of continuous-time equations of motion for the macro-particles and potentials that exactly conserves energy.  As this
method obtains the force on the macro-particles directly from the potential, we expect much lower sampling noise than when employing fields in the usual PIC algorithm.  Overall, we expect this method
to exhibit the same order of computational savings as found by Lifschitz \textit{et al.}~\cite{Lifschitz:2009aa}.  Future work will involve a non-uniform radial grid, which would be useful for
resolving the sheath in the blowout regime and is well adapted to the approach discussed.  Additionally, it is desired for computational savings to work in the moving window, which involves a trivial
transformation as discussed in Ref.~\cite{Stamm:2014aa}.  Lastly, it would be interesting to develop a symplectic integrator for the time integration of these equations of motion.

\begin{theacknowledgments}
	This work was supported by the U. S. Department of Energy under Contract No.  DE-SC0008382 and by the National Science Foundation under Contract No.  PHY-1104683.
\end{theacknowledgments}

\bibliographystyle{aipproc}  
\bibliography{AAC}

\begin{thebibliography}{9}
\expandafter\ifx\csname natexlab\endcsname\relax\def\natexlab#1{#1}\fi
\providecommand{\enquote}[1]{``#1''}
\expandafter\ifx\csname url\endcsname\relax
  \def\url#1{\texttt{#1}}\fi
\expandafter\ifx\csname urlprefix\endcsname\relax\def\urlprefix{URL }\fi
\providecommand{\eprint}[2][]{\url{#2}}

\bibitem[Huang et~al.(2006)]{Huang:2006ve}
C.~Huang, V.~K. Decyk, C.~Ren, M.~Zhou, W.~Lu, W.~B. Mori, J.~H. Cooley, T.~M.
  {Antonsen Jr.}, and T.~Katsouleas, \emph{J. Comput. Phys.} \textbf{217},
  658--679 (2006).

\bibitem[Mehrling et~al.(2014)]{Mehrling:2014aa}
T.~Mehrling, C.~Benedetti, C.~B. Schroeder, and J.~Osterhoff, \emph{Plasma
  Phys. Control. Fusion} \textbf{56}, 084012 (2014).

\bibitem[Gordon et~al.(2000)]{Gordon00}
D.~F. Gordon, W.~B. Mori, and T.~M. {Antonsen, Jr.}, \emph{{IEEE} Trans. Plasma
  Sci.} \textbf{PS-28}, 1224--1232 (2000).

\bibitem[Lifschitz et~al.(2009)]{Lifschitz:2009aa}
A.~Lifschitz, X.~Davoine, E.~Lefebvre, J.~Faure, C.~Rechatin, and V.~Malka,
  \emph{J. Comput. Phys.} \textbf{228}, 1803 -- 1814 (2009).

\bibitem[{Davidson} et~al.(2014)]{Davidson:2014aa}
A.~{Davidson}, A.~{Tableman}, W.~{An}, F.~S. {Tsung}, W.~{Lu}, J.~{Vieira},
  R.~A. {Fonseca}, L.~O. {Silva}, and W.~B. {Mori}, \emph{arXiv:1403.6890}
  (2014).

\bibitem[Evstatiev and Shadwick(2013)]{Evstatiev:2013aa}
E.~G. Evstatiev, and B.~A. Shadwick, \emph{J. Comput. Phys.} \textbf{245},
  376--398 (2013).

\bibitem[Shadwick et~al.(2014)]{Shadwick:2014aa}
B.~A. Shadwick, A.~B. Stamm, and E.~G. Evstatiev, \emph{Phys. Plasmas}
  \textbf{21}, 055708 (2014).

\bibitem[{Stamm} et~al.(2014)]{Stamm:2014aa}
A.~B. {Stamm}, B.~A. {Shadwick}, and E.~G. {Evstatiev}, \emph{IEEE Trans.
  Plasma Sci.} \textbf{42}, 1747--1758 (2014).

\bibitem[Low(1958)]{Low:1958aa}
F.~E. Low, \emph{Proc. R. Soc. London Ser. A. Math. Phys. Sci} \textbf{248},
  282--287 (1958).

\end{thebibliography}

\end{document}